\documentclass[doublecol,dvips]{epl2}
\usepackage{amsmath,amscd,amssymb,epsfig}
\usepackage{graphicx}
\usepackage{epsf}
\usepackage{pifont}

\shortauthor{}
\institute{
  \inst{1} Department of Physics and State Key Laboratory of Surface Physics, Fudan University, Shanghai 200433, China\\
  \inst{2} Department of Applied Physics, Hong Kong Polytechnic University, Hung Hom, Hong Kong,
  China\\
  \inst{*} jqyou@fudan.edu.cn
}
\abstract{We propose an approach for achieving ground-state cooling
of a nanomechanical resonator (NAMR) capacitively coupled to a
triple quantum dot (TQD). This TQD is an electronic analog of a
three-level atom in $\Lambda$ configuration which allows an electron
to enter it via lower-energy states and to exit only from a
higher-energy state. By tuning the degeneracy of the two
lower-energy states in the TQD, an electron can be trapped in a dark
state caused by destructive quantum interference between the two
tunneling pathways to the higher-energy state. Therefore,
ground-state cooling of an NAMR can be achieved when electrons
absorb readily and repeatedly energy quanta from the NAMR for
excitations.}

\begin{document}

\title{Cooling a nanomechanical resonator by a triple quantum dot}
\author{Zeng-Zhao. Li\inst{1,2} \and Shi-Hua Ouyang\inst{1,2} \and Chi-Hang
Lam\inst{2} \and J. Q. You\inst{1,*}}
\maketitle

\section{Introduction}

Nanomechanical resonators (NAMRs) with high resonance frequencies and large
quality factors are currently attracting considerable attentions owing to
their wide range of potential applications (see, e.g., \cite%
{Schwab05,Connell10}). Moreover, quantized NAMRs are potentially useful for
quantum information processing. For example, quantized motion of buckling
nanoscale bars has been proposed for qubit implementation \cite%
{SavelNJPSavel07} and also for creation of quantum entanglement \cite%
{Vitali07,Hartmann08,Tian04}. However, the dynamics of the NAMRs must
approach the quantum regime, which is usually difficult to arrive at due to
interactions with other components as well as the environments.

One way to achieve the quantum regime for an NAMR is to increase the
resonance frequency so that an energy quantum of the NAMR is larger than the
thermal energy. Recently, NAMRs based on metallic beams \cite{TF08} and
carbon nanotubes \cite{Huttel09} with resonance frequencies about several
hundred megahertz have been developed. However, a temperature lower than $10$%
~mK (below the typical dilution refrigerator temperature) is
required for an NAMR with a frequency of $200$~MHz to operate in the
quantum regime.
Therefore, 
one still needs to cool the NAMR further, for instance via coupling to an
optical or an on-chip electronic system. Various experiments on the cooling
of a single NAMR via radiation pressure or dynamical backaction have been
reported (see, e.g.,~\cite{Metzger04,Gigan06Naik06,
Kleckner06,Arcizet06,Schliesser06Kippenberg05,Poggio07,Lehnert08}). Other
cooling mechanisms based on a Cooper pair box \cite{Zhang05}
or a three-level flux qubit \cite{You08} with periodic resonant coupling
have also been theoretically proposed. In these approaches, a strong
resonant coupling between the NAMR and the qubit is required for the NAMR to
achieve its ground-state cooling.


In the weak coupling regime, a conventional method for cooling an NAMR is
the sideband cooling approach (see, e.g., ref.~\cite%
{Wilson04,Wilson07,Marquardt07,ASchliesser08,Ouyang09,Zippilli09}), where an
NAMR is usually coupled to a two-level system (TLS) with the two states
being electronic states in quantum dots \cite{Wilson04,Ouyang09,Zippilli09},
photonic states in a cavity \cite{Wilson07,Marquardt07,ASchliesser08}, or
charge states in superconducting circuits \cite{You05}. In order to achieve
ground-state cooling, the resolved-sideband cooling condition $%
\omega_m\gg\Gamma$ (with $\omega_m$ denoting the oscillating frequency of
the NAMR and $\Gamma$ the decay rate of the TLS) must be followed to
selectively drive the lowest sideband of the TLS.
However, the frequency of a typical NAMR is about $100$~MHz \cite%
{TF08,Huttel09} which is in general of the same order of the decay rate of
the two-level system, 
indicating that the resolved-sideband cooling condition is not easy to be
fully fulfilled.

\begin{figure}[tbp]
\centering
\includegraphics[width=3.4in,
bbllx=24,bblly=589,bburx=556,bbury=774]{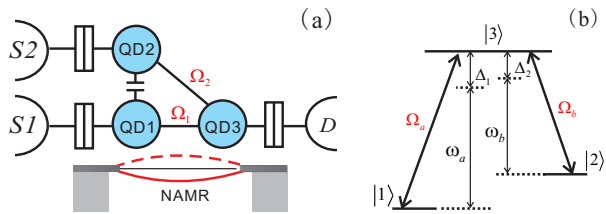}
\caption{(Color online)~(a) Schematic diagram of a TQD system. Dots $1$ and $%
2$ are both tunnel-coupled to dot $3$ (with interdot coupling strengths $%
\Omega_1$ and $\Omega_2$, respectively) while they are only capacitively
coupled to each other. An NAMR is capacitively coupled to dots $1$ and $3$
of the TQD. (b) A three-level system in $\Lambda$ configuration. The energy
detuning of the laser frequency $\protect\omega_a$ ($\protect\omega_b$) from
the transition frequency between state $|1\rangle$ ($|2\rangle$) and state $%
|3\rangle$ is $\Delta_1$ ($\Delta_2$). The corresponding driving strength of
the laser field is $\Omega_a$ ($\Omega_b$).}
\label{fig1}
\end{figure}

In this letter, we propose a different approach to cool an NAMR, in the
non-resolved sideband cooling regime, via quantum interference in a
capacitively coupled triple quantum dot (TQD) (see fig.~\ref{fig1}(a)). Here
we focus on the strong Coulomb-blockade regime with at most one electron
being allowed in the TQD at one time. The TQD acts as a three-level system
in $\Lambda$ configuration (see fig.~\ref{fig1}(b)), in which the two dot
states $|1\rangle$ and $|2\rangle$ (i.e., the single-electron orbital states
in dots 1 and 2) are coupled to a third (excited) state $|3\rangle$ via two
tunnel barriers. We will show that a dark state in the TQD can be obtained
by properly tuning the gate voltages to make the two lower-energy states
degenerate. This dark-state was previously proposed in refs.~\cite%
{Michaelis06} and \cite{Poltl09} to explain the vanishing of the charge
current transporting through a TQD system. Here, capacitively coupled to the
TQD, the NAMR can be cooled to its ground state, in analogy to the slowing
down of the atoms via quantum interference \cite{Morigi00}. Our approach has
the following potential advantages:~(i)~The cooling system presented here is
completely electronic and thus can be conveniently fabricated on a chip.
(ii)~By adjusting simply the gate voltages, the two degenerate lower-energy
states, required for destructive quantum interference, are easily achieved.
(iii)~In contrast to the cooling of an NAMR by coupling it to a
superconducting qubit \cite{Xia09} or time-dependent optical cavities\cite%
{Yong11}, the decay rate $\Gamma$ of the higher-energy state of the TQD,
which is equal to the rate of electron tunneling from the TQD to the right
electrode, is tunable by varying the gate voltage.

\section{Model}

The device layout of an NAMR coupled to a TQD is shown in fig.~\ref{fig1}%
(a). The TQD is connected to three electrodes via tunnel barriers. In the
TQD, dots $1$ and $2$ are only capacitively coupled to each other without
electrons tunneling directly between them. In contrast, electrons can tunnel
between dots $1$ and $3$ as well as between dots $2$ and $3$. Such
capacitively coupled dots have already been achieved in experiments (see,
e.g.,~\cite{Clure07}). In the strong Coulomb-blockade regime, four
electronic states need to be considered, i.e., the vacuum state $|0\rangle$,
and states $|1\rangle$, $|2\rangle$ and $|3\rangle$ corresponding to a
single electron in the respective dot. 
The NAMR is capacitively coupled to dots $1$ and $3$ as shown in fig.~\ref%
{fig1}(a). 

The total Hamiltonian of the whole system is%
\begin{equation}
H_{\mathrm{total}}\!=\!H_{0}+H_{\mathrm{int}}+H_{\mathrm{T}}+H_{\mathrm{ep}}.
\end{equation}
The unperturbed Hamiltonian $H_{0}$ is defined as $H_{0}\!=H_{\mathrm{TQD}%
}+\!H_{\mathrm{leads}}+H_{\mathrm{R}}+H_{\mathrm{ph}},$ where
\begin{eqnarray}
H_{\mathrm{TQD}}\!\!\!\!\!&=&\!\!\!\!\!-\Delta _{1}a_{1}^{\dagger
}a_{1}\!-\!\Delta _{2}a_{2}^{\dagger }a_{2}\!  \notag \\
\!\!\!\!\!&&\!\!\!\!\! + (\Omega _{1}a_{1}^{\dagger }a_{3}+\Omega
_{2}a_{2}^{\dagger }a_{3}+\mathrm{H.c.}),  \label{H-TQD}
\end{eqnarray}%
\begin{equation}
H_{\mathrm{leads}}\!=\!\sum_{i,k}E_{ik}c_{ik}^{\dagger }c_{ik},
\end{equation}%
\begin{equation}
H_{\mathrm{R}}\!=\!\omega _{m}b^{\dagger }b,
\end{equation}%
\begin{equation}
H_{\mathrm{ph}}\!=\!\sum_{q}\omega _{q}b_{q}^{\dagger }b_{q},
\end{equation}
are Hamiltonians of the TQD, the electrodes, the NAMR and the
thermal bath, respectively. We have put $\hbar =1.$ The energy of
state $|3\rangle $ is chosen as the zero-energy point and
$-\Delta_{1(2)}$ is the energy of state $|1\rangle(|2\rangle)$
relative to state $|3\rangle.$ $a_{i}^{\dagger }$ creates an
electron in the $i$th dot ($i=1,2$ or $3$) and $c_{ik}^{\dagger }$
($c_{ik}$) is the creation (annihilation) operator of an electron
with momentum $k$ in the $i$th electrode. The phonon operators
$b^{\dagger }$ and $b$ create and annihilate an excitation of
frequency $\omega _{m}$ in the
NAMR, respectively. The thermal bath is modeled as a bosonic bath with $%
b_{q}^{\dagger }$ ($b_{q}$) being the bosonic creation (annihilation)
operator at frequency $\omega _{q}$.

The electromechanical coupling between the NAMR and dots 1 and 3 of the TQD
is given by
\begin{equation}
H_{\mathrm{int}}=-(g_{3} a_{3}^{\dagger }a_{3}+g_{1} a_{1}^{\dagger
}a_{1})(b^{\dagger }+b),
\end{equation}%
with a coupling strength $g_{i}=\eta_{i} \omega _{m} (i=1, 3)$. For
simplicity we consider $g_{3}=g_{1}=g$ (or
$\eta_{3}=\eta_{1}=\eta$). For a typical electromechanical coupling,
$\eta $ $\sim 0.1$ (see, e.g., ref.~\cite{Neill09}). The tunneling
coupling between the TQD and the electrodes is
\begin{equation}
H_{\mathrm{T}}=\sum_{ik}(\Omega _{ik}\;a_{i}^{\dagger }\,c_{ik}+\mathrm{{H.c.%
}),}
\end{equation}
where $\Omega _{ik}$ characterizes the coupling strength between the $i$th
dot and the associated electrode via a tunnel barrier. Moreover, the
coupling of the NAMR to the outside thermal bath is characterized by
\begin{equation}
H_{\mathrm{ep}}=\sum_{q}\Omega _{q}(b_{q}^{\dagger }b+\mathrm{H.c.}),
\end{equation}
with $\Omega _{q}$ being the coupling strength.

\section{Quantum dynamics of the triple quantum dot}

\subsection{Analogy between a TQD and a $\Lambda$-type three-level atom}

We now show that in the absence of the NAMR, the TQD system is analogous to
a typical $\Lambda $-type three-level atom in the presence of two classical
electromagnetical fields (see fig.~\ref{fig1}(b)). The Hamiltonian of the
field-driven $\Lambda $-type three-level system can be written as $%
H_{\Lambda }\!=\!\omega _{1}a_{1}^{\dagger }a_{1}+\omega _{2}a_{2}^{\dagger
}a_{2}+\omega _{3}a_{3}^{\dagger }a_{3}+\Omega _{a}\cos (\omega
_{a}t)(a_{1}^{\dagger }a_{3}+a_{3}^{\dagger }a_{1})+\Omega _{b}\cos (\omega
_{b}t)(a_{2}^{\dagger }a_{3}+a_{3}^{\dagger }a_{2}),$ where $\omega _{i}$ ($%
i=1,2$ or $3$) is the energy of the $i$th state in the three-level system; $%
\omega _{a}$ and $\omega _{b}$ are the frequencies of the two driving fields
with $\Omega _{a}$ and $\Omega _{b}$ being the corresponding driving
strengths. In order to eliminate the time-dependence of the Hamiltonian, we
transform the system into a rotating frame defined by $U_{R}=e^{iRt}$ with $%
R=\omega _{a}a_{1}^{\dagger }a_{1}+\omega _{b}a_{2}^{\dagger }a_{2}-\omega
_{3}(a_{1}^{\dagger }a_{1}+a_{2}^{\dagger }a_{2}+a_{3}^{\dagger }a_{3}).$
Within the rotating-wave approximation, the Hamiltonian becomes
\begin{eqnarray}
\widetilde{H}_{\Lambda }\!\!\!\!\!&=&\!\!\!\!\!-\Delta _{1}a_{1}^{\dagger
}a_{1}-\Delta _{2}a_{2}^{\dagger }a_{2}  \notag \\
\!\!\!\!\!&&\!\!\!\!\!+\Omega _{1}(a_{1}^{\dagger }a_{3}+a_{3}^{\dagger
}a_{1})+\Omega _{2}(a_{2}^{\dagger }a_{3}+a_{3}^{\dagger }a_{2}),
\label{H-RWA}
\end{eqnarray}%
where $\Delta _{1}=\omega _{3}-\omega _{1}-\omega _{a}$ and $\Delta
_{2}=\omega _{3}-\omega _{2}-\omega _{b}$ are the frequency detunings while $%
\Omega _{1}=\Omega _{a}/2$ and $\Omega _{2}=\Omega _{b}/2$ are the effective
driving strengths of the two fields. It is now clear that the Hamiltonian in
eq. (\ref{H-RWA}) is formally equivalent to that of the TQD described above
(i.e., eq.~(\ref{H-TQD})).

\subsection{Dark state in the TQD}

The existence of a dark state in a $\Lambda $-type three-level atom in
quantum optics is able to suppress absorption or emission when the
lower-energy states become degenerate, i.e., $\Delta _{1}=\Delta _{2}$. We
demonstrate below that a similar dark state also exists in the TQD and can
be used for the ground-state cooling of the NAMR.

After tracing over the degrees of freedom of the electrodes, quantum
dynamics of the TQD in the absence of the NAMR is described by
\begin{eqnarray}
\dot{\rho}_{d}\!\!\!\!\!&=&\!\!\!\!\!-i[H_{\mathrm{TQD}},\rho _{d}]  \notag
\label{ME-TQD} \\
\!\!\!\!\!&&\!\!\!\!\!+\Gamma _{1}\mathcal{D}[a_{1}^{\dagger }]\rho +\Gamma
_{2}\mathcal{D}[a_{2}^{\dagger }]\rho _{d}+\Gamma _{3}\mathcal{D}[a_{3}]\rho
_{d},
\end{eqnarray}%
where $\rho _{d}$ is the reduced density matrix of the TQD and $\Gamma _{i}$
($i=1,~2$ or $3$) is the rate for electrons tunneling into or out of the $i$%
th dot. The notation $\mathcal{D}$ for any operator $A$ is given by $%
\mathcal{D}[A]\rho =A\rho A^{\dagger }-\frac{1}{2}(A^{\dagger }A\rho +\rho
A^{\dagger }A).$ Considering equal energy detunings of the two lower-energy
states $|1\rangle $ and $|2\rangle $ with respect to the excited state $%
|3\rangle $, i.e., $\Delta _{1}=\Delta _{2}=\Delta $, the eigenstates $%
|g\rangle $, $|-\rangle $ and $|+\rangle $ of the TQD become $|g\rangle
\!=\!\beta |3\rangle -\frac{\alpha }{\Omega }(\Omega _{1}|1\rangle +\Omega
_{2}|2\rangle ),$ $|-\rangle \!=\!\frac{1}{\Omega }(\Omega _{2}|1\rangle
-\Omega _{1}|2\rangle )$ and $|+\rangle \!=\!\alpha |3\rangle +\frac{\beta }{%
\Omega }(\Omega _{1}|1\rangle +\Omega _{2}|2\rangle ),$ where $\alpha =\cos
(\theta /2)$, $\beta =\sin (\theta /2)$, $\tan \theta =2\Omega /\Delta $ and
$\Omega =\sqrt{\Omega _{1}^{2}+\Omega _{2}^{2}}$. The corresponding
eigenenergies are $E_{\mathrm{g}}=-\frac{\Delta +\phi }{2},~~E_{-}=-\Delta$
and $E_{+}=-\frac{\Delta -\phi }{2},$ with $\phi =\sqrt{\Delta ^{2}+4\Omega
^{2}}$. For simplicity, we consider equal couplings of the three quantum
dots to the corresponding electrodes, i.e., $\Gamma _{1}=\Gamma _{2}=\Gamma
_{3}\equiv \Gamma $. Based on the eigenstate basis of the TQD, the equations
of motion for the reduced density matrix elements $\rho_{ij}\equiv\langle
i|\rho_{d}|j \rangle$ ($i, j = 0, g, -$ and $+$) of the TQD are obtained
from eq.~(\ref{ME-TQD}) as
\begin{eqnarray}
\dot{\rho}_{00}\!\!&\!\!=\!\!&\!\!-2\Gamma \rho _{00}+\Gamma \beta ^{2}\rho
_{gg}+\Gamma \alpha ^{2}\rho _{++}+\Gamma \alpha \beta (\rho _{+g}+\rho
_{g+}),  \notag \\
\dot{\rho}_{gg}\!\!&\!\!=\!\!&\!\!\Gamma \alpha ^{2}\rho _{00}-\Gamma \beta
^{2}\rho _{gg}-\frac{\Gamma }{2}\alpha \beta (\rho _{+g}+\rho _{g+}),  \notag
\\
\dot{\rho}_{--}\!\!&\!\!=\!\!&\!\!\Gamma \rho _{00},  \notag \\
\dot{\rho}_{++}\!\!&\!\!=\!\!&\!\!\Gamma \beta ^{2}\rho _{00}-\Gamma \alpha
^{2}\rho _{++}-\frac{\Gamma }{2}\alpha \beta (\rho _{+g}+\rho _{g+}),  \notag
\\
\dot{\rho}_{+g}\!\!&\!\!=\!\!&\!\!-i(E_{\mathrm{+}}-E_{\mathrm{g}})\rho
_{+g}-\frac{\Gamma }{2}\rho _{+g}-\Gamma \alpha \beta \rho _{00}  \notag \\
&&-\frac{\Gamma }{2}\alpha \beta (\rho _{++}+\rho _{gg}) .  \label{DME}
\end{eqnarray}
Here the diagonal elements $\rho_{00},$ $\rho_{gg},$ $\rho_{--}$ and
$\rho_{++}$ are the probabilities of TQD in the states $|0\rangle,$
$|g\rangle,$ $|-\rangle$ and $|+\rangle,$ respectively, while the
off-diagonal element $\rho_{+g}$ involves the quantum coherence
between states $|+\rangle$ and $|g\rangle.$ Effective electron
tunneling processes through the TQD are described by these equations
above. Firstly, an electron can tunnel from the left electrodes into
any of the three eigenstates $|g\rangle $, $|-\rangle $ and
$|+\rangle $
of an initially empty TQD, with tunneling rates $\Gamma \alpha ^{2}$, $%
\Gamma $ and $\Gamma \beta ^{2},$ respectively (see
eq.~(\ref{DME})). Note that the total tunneling rate is $2\Gamma $
since each of the two tunnel barriers admits electrons with rate
$\Gamma$. Then, an electron in the eigenstate $|g\rangle $
($|+\rangle $) can move out of the TQD to the right electrode with a
rate $\Gamma \beta ^{2}$ ($\Gamma \alpha ^{2}$) (see
eq.~(\ref{DME})). However, no further tunneling occurs if the state $%
|-\rangle $, which is orthogonal to $|3\rangle $, is occupied. This results
from the destructive quantum interference between the transition $|1\rangle
\rightarrow |3\rangle $ (i.e., the electron tunneling from state $|1\rangle $
to state $|3\rangle $ in the TQD system) and the transition $|2\rangle
\rightarrow |3\rangle $ (i.e., the electron tunneling from $|2\rangle $ to $%
|3\rangle $). Therefore, an electron can be trapped in the state $|-\rangle $
which is called a dark state \cite{Michaelis06}. Furthermore, this dark
state depends on the initial condition and the probability of having no
electron in the TQD (see eq.~(\ref{DME})). Luckily, within the time scale
for the ground-state cooling of the NAMR, the dark state will always arise
as shown below.

\section{Ground-state cooling of the nanomechanical resonator}

\subsection{Quantum dynamics of coupled NAMR-TQD system}

Rather than analyzing directly the energy exchange between the NAMR and the
TQD which involves tedious algebra, we apply a canonical transform $U=e^{S}$
on the whole system, where $S=\eta (a_{3}^{\dagger }a_{3}+a_{1}^{\dagger
}a_{1})(b-b^{\dagger }).$ The transformed Hamiltonian is given by
\begin{eqnarray}
H\!\!\!\!\! &=&\!\!\!\!\!UH_{\mathrm{total}}U^{\dagger }  \notag \\
\!\!\!\!\! &=&\!\!\!\!\!H_{\mathrm{leads}}+H_{\mathrm{ph}}+\omega
_{m}b^{\dagger }b-\Delta _{1}a_{1}^{\dagger }a_{1}-\Delta _{2}a_{2}^{\dagger
}a_{2}  \notag \\
\!\!\!\!\! &&\!\!\!\!\!+(\Omega _{1}a_{1}^{\dagger }a_{3}+\Omega
_{2}a_{2}^{\dagger }a_{3}B+\mathrm{H.c.})  \notag \\
\!\!\!\!\! &&\!\!\!\!\!+\sum_{q}\{\Omega _{q}\;b_{q}^{\dagger }\,[b+\eta
(a_{3}^{\dagger }a_{3}+a_{1}^{\dagger }a_{1})]+\mathrm{H.c.}\}  \notag \\
\!\!\!\!\! &&\!\!\!\!\!+\sum_{k}[\Omega _{1k}\;a_{1}^{\dagger
}\,c_{1k}B^{\dagger }+\Omega _{2k}\;a_{2}^{\dagger }\,c_{2k}  \notag \\
\!\!\!\!\! &&\!\!\!\!\!+\Omega _{3k}\;a_{3}^{\dagger }\,c_{3k}B^{\dagger }+%
\mathrm{{H.c.}]},  \label{H-Us}
\end{eqnarray}%
where we have neglected a small level shift of $\eta ^{2}\omega _{m}$ to the
states $|1\rangle $ and $|3\rangle $ and defined $B=\exp [{-\eta
(b-b^{\dagger })}].$ To describe the quantum dynamics of the coupled
NAMR-TQD system, we have derived a master equation (under the Born-Markov
approximation) by tracing over the degrees of freedom of both the electrodes
and the thermal bath. Up to second order in $\eta $, the master equation can
be written as
\begin{eqnarray}
\frac{d\rho }{dt}\!\!\!\!\! &=&\!\!\!\!\!-i\omega _{m}[b^{\dagger }b,\rho
]-i[H_{\mathrm{TQD}},\rho ]-i[V(b^{\dagger }-b),\rho ]  \notag \\
\!\!\!\!\! &&\!\!\!\!\!-i[V^{\prime }(b^{\dagger }-b)^{2},\rho ]+\mathcal{L}%
_{\mathrm{T}}\rho +\mathcal{L}_{\mathrm{D}}\rho ,  \label{ME}
\end{eqnarray}%
where
\begin{eqnarray}
V\!\!\!\!\! &=&\!\!\!\!\!\eta \Omega _{2}(a_{2}^{\dagger
}a_{3}-a_{3}^{\dagger }a_{2}),V^{\prime }\!=\!\frac{\eta ^{2}}{2}\Omega
_{2}(a_{2}^{\dagger }a_{3}+a_{3}^{\dagger }a_{2}),  \notag \\
\mathcal{L}_{\mathrm{T}}\rho \!\!\!\!\! &=&\!\!\!\!\!\mathcal{L}_{\mathrm{T}%
}^{\left( 1\right) }\rho +\mathcal{L}_{\mathrm{T}}^{\left( 2\right) }\rho ,\!
\notag \\
\mathcal{L}_{\mathrm{T}}^{\left( 1\right) }\rho \!\!\!\!\!
&=&\!\!\!\!\!\Gamma \mathcal{D}[a_{1}^{\dagger }]\rho +\Gamma \mathcal{D}%
[a_{2}^{\dagger }]\rho +\Gamma \mathcal{D}[a_{3}]\rho ,  \notag \\
\mathcal{L}_{\mathrm{T}}^{\left( 2\right) }\rho \!\!\!\!\!
&=&\!\!\!\!\!-\eta ^{2}\Gamma \mathcal{D}[a_{1}^{\dagger }]\rho -\eta
^{2}\Gamma \mathcal{D}[a_{3}]\rho   \notag \\
\!\!\!\!\! &&\!\!\!\!\!+\eta ^{2}\Gamma (\mathcal{D}[a_{1}^{\dagger
}b^{\dagger }]\rho +\mathcal{D}[a_{1}^{\dagger }b]\rho   \label{Ld} \\
\!\!\!\!\! &&\!\!\!\!\!+\mathcal{D}[a_{3}b^{\dagger }]\rho +\mathcal{D}%
[a_{3}b]\rho )  \notag \\
\!\!\!\!\! &&\!\!\!\!\!+\eta ^{2}\Gamma (a_{1}a_{1}^{\dagger }b^{\dagger
}b\rho +\rho b^{\dagger }ba_{1}a_{1}^{\dagger }-a_{1}^{\dagger }\rho
b^{\dagger }ba_{1}  \notag \\
\!\!\!\!\! &&\!\!\!\!\!-a_{1}^{\dagger }b^{\dagger }b\rho
a_{1}+a_{3}^{\dagger }b^{\dagger }ba_{3}\rho +\rho a_{3}^{\dagger
}b^{\dagger }ba_{3}  \notag \\
\!\!\!\!\! &&\!\!\!\!\!-a_{3}\rho a_{3}^{\dagger }b^{\dagger }b-b^{\dagger
}ba_{3}\rho a_{3}^{\dagger }),  \notag \\
\mathcal{L}_{\mathrm{D}}\rho \!\!\!\! &=&\!\!\!\!\gamma \lbrack n(\omega
_{m})+1]\mathcal{D}[b]\rho +\gamma n(\omega _{m})\mathcal{D}[b^{\dagger
}]\rho .  \notag
\end{eqnarray}%
Here $\mathcal{L}_{\mathrm{T}}$ and $\mathcal{L}_{\mathrm{D}}$ are
Liouvillian operators: $\mathcal{L}_{\mathrm{T}}\rho $ accounts for the
dissipation due to the electrodes and $\mathcal{L}_{\mathrm{D}}\rho $ the
dissipation in the NAMR induced by the thermal bath. Also, $\gamma $ denotes
the decay rate of excitations in the NAMR induced by the thermal bath and $%
n(\omega _{m})$ is the average boson number at frequency $\omega
_{m}$ in the thermal bath. Usually, $\gamma \ll \Gamma$ for an NAMR,
so $\mathcal{L}_{\mathrm{D}}\rho$ is a higher-order term.

\subsection{Quantum dynamics of the NAMR}

In order to extract the dynamics of NAMR, we trace out the degrees of
freedom of the TQD, which can be regarded as part of the environment
experienced by the NAMR. This is achieved through adiabatical elimination
\cite{Wilson07,Cirac92} in the limit $\gamma \ll g\ll \omega _{m},$
corresponding to a weak coupling of the NAMR with the TQD. At zeroth order
in $\eta$ in which NAMR and TQD are decoupled, the dynamics of the whole
system is governed by
\begin{equation}
\frac{d\rho }{dt}=\mathcal{L}_{\mathrm{0}}\rho,  \label{Zeroth-ME}
\end{equation}
where the Liouvillian operator $\!\!\mathcal{L}_{\mathrm{0}}\!\!$ is
defined by $\!\mathcal{L}_{\mathrm{0}}\rho\!\!\equiv \!\!-\!i\omega
_{m}[b^{\dagger }b,\rho ]\!-\!i[H_{\mathrm{TQD}},\rho
]+\mathcal{L}_{\mathrm{T}}^{\left( 1\right) }\rho$. The steady-state
solution to eq.~(\ref{Zeroth-ME}) can be expanded
\cite{Cirac92,Morigi00} in the basis of $\rho _{nn^{\prime
}}\!=\!\vert n\rangle \langle n^{\prime}\vert \otimes \rho_{ss},$
with $\vert n\rangle$ being eigenstates of the NAMR Hamiltonian
(i.e., $ H_{R}\left\vert n\right\rangle \!=\!n \omega _{m}\left\vert
n\right\rangle \!$) and $\rho_{ss}$ the density operator $\rho_{d}$
of the TQD in the steady state (which is the solution to
eq.~(\ref{ME-TQD}) when $d\rho_{d}/dt=0$). Here $\rho_{nn^{\prime}}$
are the eigenvectors of $\mathcal{L}_{\mathrm{0}}$:
$\mathcal{L}_{\mathrm{0}}\left\vert n\right\rangle \!\left\langle
n^{\prime }\right\vert \otimes \rho _{ss}\!=0$ $\left( n, n^{\prime
}=0,1,\ldots \right),$ corresponding to the eigenvalues $\lambda
_{nn^{\prime }}\!=\!-i\left( n-n^{\prime }\right) \omega _{m}.$ In
particular, for the zero eigenvalue $\lambda_{0} \equiv
\lambda_{nn}=0,$ the eigenvectors are given by
$\mathcal{L}_{\mathrm{0}}\left\vert n\right\rangle \!\left\langle
n\right\vert \otimes \rho _{ss}\!=\!0$ $\left( n=0,1,\ldots
\right),$ which are infinitely degenerate. The zeroth-order
Liouville eigenstates with $\mathcal{\lambda }_{\mathrm{0}}=0$ are
connected
by the operators $\mathcal{L}_{\mathrm{1 }}$ ($\mathcal{L}_{\mathrm{1}%
}\rho\!\equiv\!-i[V(b^{\dagger }-b),\rho ]$)
and $\mathcal{L}_{\mathrm{2}}$ \!($\mathcal{L}_{\mathrm{2}%
}\rho\!\equiv\!-i[V^{\prime}(b^{\dagger }-b)^2,\rho ]+\mathcal{L}_{\mathrm{T}%
}^{\left( 2\right) }\rho+\mathcal{L}_{\mathrm{D}}\rho $)
to the subspaces associated with nonzero eigenvalues $\mathcal{\lambda }_{%
\mathrm{k}}\neq 0$ $(k=1,2,\cdots).$ This is due to the coupling between the
NAMR and TQD. In the regime $g\ll \omega _{m}$ $\left(i.e., \eta \ll
1\right) $, such coupling
is weak and can be analyzed using perturbation theory and adiabatic
elimination. We define a projection operator $\!\mathcal{P}$ on the subspace
with zero eigenvalue $\!\mathcal{\lambda }_{\mathrm{0}}=0$ of $\mathcal{L}_{%
\mathrm{0}}$ according to%
\begin{equation}
\mathcal{P}\rho \!\equiv \mathcal{P}^{NAMR}\otimes \mathcal{P}^{TQD},
\end{equation}%
where 
\begin{eqnarray}
\mathcal{P}^{NAMR}X \!\!&\!\!\equiv \!\!&\!\!\sum_{n}\left\vert
n\right\rangle \!\left\langle n\right\vert \!\left\langle
n\right\vert
X\left\vert n\right\rangle ,  \notag \\
\mathcal{P}^{TQD}X \!\!&\!\!\equiv \!\!&\!\!\rho _{ss}Tr_{TQD}X.
\end{eqnarray}%
Projection of the above master equation~(\ref{ME}) gives the master
equation for the reduced density matrix $\mu$ of the NAMR, up to
second order in $\eta $,
\begin{eqnarray}
\dot{\mu}\!\!\!\!\!&=&\!\!\!\!\!-i(\omega _{m}+\delta _{m})[b^{\dagger
}b,\,\mu ]+\frac{1}{2}\left\{ \gamma \lbrack n(\omega _{m})+1]+A_{-}(\omega
_{m})\right\}  \notag \\
\!\!\!\!\!&&\!\!\!\!\!\times \lbrack 2b\mu b^{\dagger }-(b^{\dagger }b\mu
+\mu b^{\dagger }b)]  \notag \\
\!\!\!\!\!&&\!\!\!\!\!+\frac{1}{2}[\gamma n(\omega _{m})+A_{+}(\omega
_{m})][2b^{\dagger }\mu b-(bb^{\dagger }\mu +\mu bb^{\dagger })].
\label{ME-MR}
\end{eqnarray}%
Here $\delta _{m}$ is the driving-induced shift of the NAMR frequency given
by $\delta _{m}\!\!=\!\!\mathop{\rm Im}\left[ G\left( i\omega _{m}\right)
+G\left( -i\omega _{m}\right) \right],$ where $G\left( s\right)
\!=\!-\langle \widetilde{V}\left( s\right) V\left( 0\right) \rangle\!$ with $%
\widetilde{V}\left( s\right)$ being the Laplace transform of $V\left(
s\right).$ 
In eq.~(\ref{ME-MR}), $A_{+}$ and $A_{-}$ are induced by the coupling to the
TQD and are given by $A_{\pm }=2\mathop{\rm Re}\left[ G\left( \pm i\omega
_{m}\right) \right] +\eta ^{2}\Gamma \left( \rho _{00}^{\mathrm{st}}+\rho
_{33}^{\mathrm{st}}\right),$ where $\rho _{00}^{\mathrm{st}}$ and $\rho
_{33}^{\mathrm{st}}$ are steady-state probabilities of the states $|0\rangle$
(empty TQD) and $|3\rangle$ (single electron in dot 3), respectively.

\subsection{Average phonon number}

From master equation~(\ref{ME-MR}), the evolution equation for the phonon
number probability distribution $p_{n}=\langle n|\mu |n\rangle $ of the NAMR
is obtained as
\begin{eqnarray}
\frac{dp_{n}}{dt} \!\!\!\!\!&=&\!\!\!\!\!\big\{\gamma \lbrack n(\omega
_{m})+1]+A_{-}\big\}\lbrack (n+1)p_{n+1}-np_{n}]  \notag \\
\!\!\!\!\!&&\!\!\!\!\!+[\gamma n(\omega _{m})+A_{+}][np_{n-1}-(n+1)p_{n}].
\label{pn}
\end{eqnarray}%
Hence the evolution of the average phonon number, $\langle n\rangle
=\sum_{n}np_{n}$, in the NAMR is described by
\begin{equation}
\frac{d\left\langle {n}\right\rangle }{dt}=-(\gamma +W)\langle n\rangle +{%
\gamma n(\omega _{m})+A_{+}},  \label{average-phn}
\end{equation}%
where $W=A_{-}-A_{+}$. In order to cool the NAMR, one needs $W>0$ (i.e., $%
A_{-}>A_{+}$). Consequently, the steady-state average phonon number in the
NAMR is
\begin{equation}
{n}_{\mathrm{st}}=\frac{\gamma n(\omega_m)+A_+}{\gamma+W}.  \label{phn}
\end{equation}
Here the term $\gamma n(\omega_m)$ in the numerator is due to the
thermal bath while $A_+$ results from the scattering processes by
the TQD. We assume that the NAMR is initially at equilibrium with
the thermal bath, so that its phonon number is initially
$n(\omega_m)$. In order to cool down the NAMR significantly, i.e.,
$n_{\mathrm{st}} \ll n(\omega_m)$, one needs a large cooling rate
$W\gg\gamma.$ This indeed can be achieved using typical experimental
parameters and will be shown below.

The transition rates $A_{\pm }$ under the condition $\Delta _{1}=\Delta
_{2}\equiv \Delta $ appropriate for the dark state are found to be
\begin{eqnarray}
A_{\pm } \!\!\!\!\!&=&\!\!\!\!\!\frac{4\eta ^{2}\Omega _{1}^{2}\Omega
_{2}^{2}}{\Omega ^{2}}\frac{\omega _{m}^{2}\Gamma }{4[\Omega ^{2}-\omega
_{m}(\omega _{m}\pm \Delta )]^{2}+\omega _{m}^{2}\Gamma ^{2}}  \notag \\
\!\!\!\!\!&&\!\!\!\!\!+\eta ^{2}\Gamma \left( \rho _{00}^{\mathrm{st}}+\rho
_{33}^{\mathrm{st}}\right).  \label{Apm}
\end{eqnarray}%
To cool the NAMR, one needs $A_{-}>A_{+}$, which is fulfilled either when $%
\Delta >0$ and $\Omega <\omega _{m}$, or when $\Delta <0$ and $\Omega
>\omega _{m}$. Assuming also $W\gg \gamma $, the steady-state average phonon
number in the NAMR is approximately given by ${n}_{\mathrm{st}}\approx {%
\gamma n(\omega _{m})}/{W}+n_{f},$ where
\begin{equation}
n_{f}\equiv \frac{A_{+}}{W}=\frac{4[\Omega ^{2}-\omega _{m}(\omega
_{m}-\Delta )]^{2}+\omega _{m}^{2}\Gamma ^{2}}{16\Delta \omega _{m}(\omega
_{m}^{2}-\Omega ^{2})}.  \label{nf}
\end{equation}
It is easy to see that $n_f$ reaches a minimum $n_f^{\mathrm{min}%
}\!\!=\!\!\left({\Gamma}/{4\Delta}\right)^2,$ when the term inside the
square brackets in the r.h.s. of eq.~(\ref{nf}) becomes zero, i.e.,
\begin{equation}
\Omega^2=\omega_m(\omega_m-\Delta),  \label{condition}
\end{equation}
or $\omega_m=(\phi+\Delta)/2.$ The optimal cooling condition in eq.~(\ref%
{condition}) can be fulfilled by properly choosing the parameters $%
\Omega,~\omega_m$, and $\Delta$. Then, the steady-state average phonon
number in the NAMR can be much smaller than unity provided $\Delta\gg\Gamma$%
, implying that ground-state cooling of the NAMR is possible. The phonon
number $n_f$ attainable according to eq.~(\ref{nf}) is identical to the
previous result for the cooling of trapped atoms via quantum interference
\cite{Morigi00}. However, solid-state cooling system proposed here has
notable advantages such as easy fabrication on a single chip and high
controllability. Specifically, all the relevant parameters (i.e., the
detuning $\Delta$, the tunneling rate $\Gamma$ and the interdot coupling
strengths $\Omega_1$ and $\Omega_2$) can be controlled by tuning the gate
voltages in the TQD. Thus, the optimal cooling condition in eq.~(\ref%
{condition}) can be conveniently fulfilled for a specified frequency $%
\omega_m$ of the NAMR.

\begin{figure}[tbp]
\centering
\includegraphics[scale=0.75]{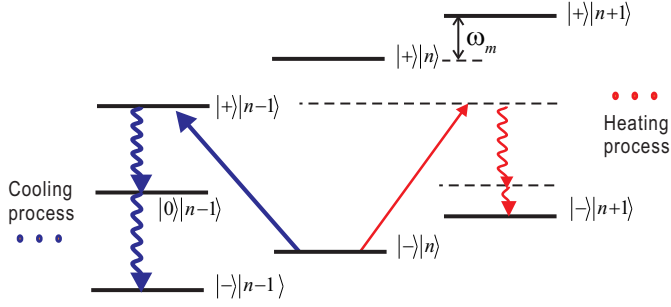}
\caption{~(Color online) Schematic diagram of the transitions in the coupled
NAMR-TQD system, where the cooling process dominates over the heating
process. When the frequency of the NAMR is equal to the transition frequency
between states $|-\rangle$ and $|+\rangle,$ i.e., $\protect\omega_m=(\protect%
\phi+\Delta)/2,$ the process $|-,n\rangle\rightarrow$ $|+,n-1\rangle$ is
resonantly enhanced while the process $|-,n\rangle\rightarrow$ $|+,n+1\rangle
$ is suppressed. A subsequent tunneling of an electron from the excited
state $|+\rangle$ to the right electrode (i.e., $|+,n-1\rangle\rightarrow$ $%
|0,n-1\rangle$) is followed by another tunneling process with an electron
transporting from the left electrode to the dark state $|-\rangle$ (i.e., $%
|0,n-1\rangle\rightarrow$ $|-,n-1\rangle$), so as to extract a quantum of
energy from the NAMR.}
\label{fig2}
\end{figure}

\begin{figure}[tbp]
\centering
\includegraphics[width=2.8in,
bbllx=20,bblly=17,bburx=518,bbury=390]{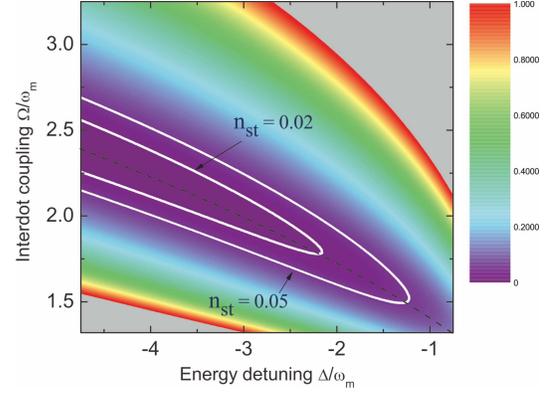}
\caption{~(Color online) Contour plot of the steady-state average phonon
number ${n}_{\mathrm{st}}$ in the NAMR as a function of the normalized
driving detuning $\Delta/\protect\omega_m$ and the interdot coupling $\Omega/%
\protect\omega_m$. The two solid curves correspond to ${n}_{\mathrm{st}}=0.05
$ and $0.02$. The black dashed line represents $\Omega^2=\protect\omega_m(%
\protect\omega_m-\Delta)$, under which the NAMR can be optimally cooled. We
have chosen $\Omega_1=\Omega_2=\Omega/\protect\sqrt{2}$ and typical
parameters $\protect\omega_m=2\protect\pi\times100$~MHz, $\Gamma=\protect%
\omega_m$, $Q=10^5$, $\protect\eta=0.1$, and $n(\protect\omega_m)=21$.}
\label{fig3}
\end{figure}

The underlying physics of the optimal cooling condition in eq.~(\ref%
{condition}) can be understood more intuitively in the eigenstate basis of
the TQD. In the limit $\gamma\ll g\ll\omega_m$ considered here, the TQD
arrives quickly at the dark state $|-\rangle,$ compared with the dynamic
time scale of the NAMR. Therefore, the TQD is practically always maintained
in the dark state. The coupling between the NAMR and the TQD excites the
electron to the state $|+\rangle$ most readily when the frequency $\omega_m$
of the NAMR is equal to the transition frequency $(\phi+\Delta)/2$ between
the states $|-\rangle$ and $|+\rangle$, i.e., $\omega_m=(\phi+\Delta)/2$.
This corresponds to the transition $|-,n\rangle\rightarrow|+,n-1\rangle$.
The excited electron subsequently tunnels to the right electrode, i.e., $%
|+,n-1\rangle\rightarrow|0,n-1\rangle.$ Meanwhile, another electron tunnels
into the TQD, and then the dark state returns promptly, i.e., $%
|0,n-1\rangle\rightarrow|-,n-1\rangle.$ This whole process extracts an
energy quantum from the NAMR. Therefore, the NAMR will be cooled to the
ground state when this cycle repeats, i.e., $|-,n\rangle\rightarrow$ $%
|+,n-1\rangle\rightarrow$ $|0,n-1\rangle\rightarrow$ $|-,n-1\rangle%
\rightarrow$$\cdots$ , as illustrated in fig.~\ref{fig2}. Here we emphasize
that the condition of resonance transition for TQD from the state $|-\rangle$
to the state $|+\rangle$ via the NAMR is equivalent to the optimal cooling
condition in eq.~(\ref{condition}). Although an electron may also relax from
the dark state $|-\rangle$ to the ground state $|g\rangle$ by releasing
energy to the NAMR, this heating process of the NAMR is strongly suppressed
because the frequency of the NAMR is off-resonant to the transition $%
|-\rangle\rightarrow|g\rangle$ in the TQD.

Figure \ref{fig3} displays a contour plot of the steady-state average phonon
number of the NAMR (${n}_{\mathrm{st}}$) as a function of the effective
interdot coupling $\Omega$ ($=\sqrt{\Omega_1^2+\Omega_2^2}$) and the energy
detuning $\Delta$. Here we choose $\Delta<0$ and $\Omega>\omega_m$ to make
sure that $W>0$. For these typical parameters, a small ${n}_{\mathrm{st}%
}<0.05$ is predicted over a wide range of values on the
$\Omega-\Delta$ plane, which implies that ground-state cooling of
the NAMR should be experimentally accessible. Furthermore, from
eq.~(\ref{Apm}), we obtain a cooling rate
$W\approx2\pi\times8.52$~MHz by using typical
experimental parameters \cite{TF08,Onac06,Gustavsson07}: $%
\omega_m\!=2\pi\times100$~MHz, $\Delta=-2\pi\times40$~GHz and $%
g=2\pi\times10 $~MHz, as well as by choosing the interdot couplings %
$\Omega_1=\Omega_2\simeq2\pi\times1.41$~GHz to fulfill the optimal
cooling condition $\Omega^2=\omega_m(\omega_m-\Delta)$. Considering
an NAMR with a quality factor $Q=10^5$ (see, e.g.,
ref.~\cite{Huttel09}), one has $\gamma=\omega_m/Q=2\pi\times1$kHz.
Therefore, appreciable cooling with $W\gg\gamma$ can be achieved. In
this case, an NAMR can be cooled from,
e.g., an initial temperature $T_{m}=100$~mK, corresponding to $%
n(\omega_m)=21,$ down to $T=0.8$~mK with ${n}_{\mathrm{st}}=0.0025$.
Starting from an initial temperature $T_{m}$, the final temperature
of the cooled NAMR should be bound by \cite{Grajcar08}
$T^{\ast}\!=\frac{\omega _{m}}{|\Delta| }T_{m}$. In the parameter
regime we studied above, $T^{\ast }=0.25$~mK, which is indeed lower
than the achieved temperature $T=0.8$~mK. Moreover, it should be
noted that the resolved-sideband cooling condition
$\omega_m\gg\Gamma$ is not required in our scheme. Finally, note
that in addition to the electrodes coupled to the TQD, there are
incoherent processes induced by other degrees of freedom (e.g.,
background charge fluctuations) in the environment. These processes
will affect the dark state of the TQD and then limit the cooling
efficiency. Thus, one needs to produce a high-quality TQD with good
quantum coherence.


\section{Conclusion}

In summary, we have proposed an approach for achieving the ground-state
cooling of an NAMR. It is shown that a dark state, which is decoupled from
the excited state, can appear in the TQD in the absence of the NAMR when its
two lower-energy localized states become degenerate. With the NAMR
capacitively coupled to the TQD and being in resonance with the transition
between the dark state and the excited eigenstate in the TQD, we have shown
that the ground-state cooling of an NAMR can be achieved in the non-resolved
sideband cooling regime.

\acknowledgments This work was supported by the National Basic Research
Program of China Grant No. 2009CB929300, the National Natural Science
Foundation of China Grant No. 10625416, and the Hong Kong GRF Grant No.
5009/08P.

\end{document}